\documentclass[twocolumn]{revtex4}
\usepackage{amsmath}
\usepackage{amsfonts}
\usepackage{amssymb}
\usepackage{graphicx}
\usepackage{placeins}
\usepackage{float}
\usepackage{subfigure}
\usepackage{subfigure}
\usepackage{color}
\input{Qcircuit}
\begin{document}
\title{Entanglement in the Grover's Search Algorithm}
\author{Shantanav Chakraborty}
\email{shantanav_with_u@iitj.ac.in}
\author{Subhashish Banerjee}
\email{subhashish@iitj.ac.in}
\author{Satyabrata Adhikari}
\email{satya@iitj.ac.in}
\author{Atul Kumar}
\email{akumar@iitj.ac.in}
\affiliation{Indian Institute of Technology Jodhpur, Jodhpur-342011, India}
\begin{abstract}
Quantum Algorithms have long captured the imagination of computer scientists and physicists primarily because of the speed up achieved by them over their classical counterparts using principles of quantum mechanics. Entanglement is believed to be the primary phenomena behind this speed up. However their precise role in quantum algorithms is yet unclear. In this article, we explore the nature of entanglement in the Grover's search algorithm. This algorithm enables searching of elements from an unstructured database quadratically faster than the best known classical algorithm. Geometric measure of entanglement has been used to quantify and analyse entanglement across iterations of the algorithm. We reveal how the entanglement varies with increase in the number of qubits and also with the number of marked or solution states. Numerically, it is seen that the behaviour of the maximum value of entanglement is monotonous with the number of qubits. Also, for a given value of the number of qubits, a change in the marked states alters the amount of entanglement. The amount of entanglement in the final state of the algorithm has been shown to depend solely on the nature of the marked states. Explicit analytical expressions are given showing the variation of entanglement with the number of iterations and the global maximum value of entanglement attained across all iterations of the algorithm. 
\end{abstract}
\maketitle
\begin{section}{Introduction}
Entanglement is a purely quantum mechanical phenomena which lies at the heart of many tasks of quantum information and quantum computation \cite{NieChu01}. Entanglement is perceived as a resource which facilitates faster and more secured communication as compared to classical means \cite{Bru02}. It is believed to be the primary reason behind the speed up achieved by quantum algorithms over their classical counterparts \cite{Delgado1}. However, the lack of a mathematical structure for higher qubits make the study of entanglement difficult \cite{Aaron04}.
At the heart of quantum algorithm, lies two fundamental algorithms namely Shor's algorithm \cite{shor94} and the Grover's Search Algorithm \cite{Gro96}. The Shor's algorithm, developed by Peter Shor in 1994, factors a number into primes in polynomial time. Now this algorithm, when implemented in a quantum computer, poses a risk for the existing crypto-system. It has been shown that entanglement is necessary to achieve an exponential speed up in Shor's Algorithm \cite{DJ92}. In \cite{Orus}, scaling of entanglement was studied in Shor's algorithm and adiabatic quantum algorithms across quantum phase transitions in Grover's algorithm as well as the NP-Complete Exact Cover problem. It was observed that the scaling of entanglement defined the complexity of the quantum phase transitions. \\
In this article, the focus would be on the Grover's search algorithm. The problem of searching for an entry in an unordered database requires $O(N)$ time even for the best known classical randomized algorithm. The Grover's algorithm achieves this task in $O(\sqrt{N})$ time. In fact it has been shown that there exists a family of quantum search algorithms that can achieve a quadratic speedup of which the Grover's algorithm is a distinguished member \cite{Delgado2}. Also, the quantum search algorithm would require an exponential overhead in terms of resources, if implemented without entanglement \cite{Meyer1}. \\
In \cite{Shan13}, the non-classical correlations for the two qubit scenario of the quantum search algorithm had been studied, however, in this article we consider the general $n$ qubit and $M$ solution states scenario. For an $n$ qubit system, the algorithm searches for $M$ elements that are stored in a database of $N=2^n$ elements. The algorithm creates an initial superposition of states on applying Hadamard gates to each of the $n$ qubits resulting in an equal superposition of all the basis states. This is followed by the repeated application of the Grover iterate $G$ thereby amplifying the amplitudes of the marked states. The circuit for the Grover's algorithm is shown in Fig. \ref{grocircuit}.   
\begin{figure}[h]
\centering
\begin{align*}
 \Qcircuit @C=1em @R=.7em {   
  &						&								&				 & & &\mbox{\textbf{Repeat $O(\sqrt{N/M})$ times}} & & \\
  & \lstick{|0\rangle} & /^n \qw & \gate{H^{\otimes n}} & \multigate{1}{U} & \gate{H^{\otimes n}} & \gate{2 |{0^n}\rangle \langle{0^n}| - I_n}         & \gate{H^{\otimes n}} & \qw  \\
 & \lstick{|1\rangle} & \qw     & \gate{H}     & \ghost{U}        & \qw                  & \qw     &\qw   &\qw \qw \gategroup{2}{5}{3}{8}{.7em}{--} \\                              
 }
 \end{align*}
 \caption{Circuit for Grover's Algorithm}
\label{grocircuit}
\end{figure}
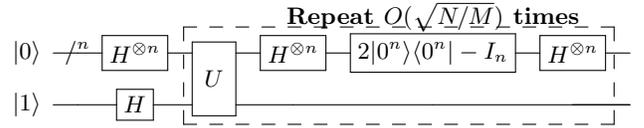 
\\
The repeated application of $G$ results in the rotation of initial superposition of an n-qubit product state $|\psi_0\rangle$ towards the $M$ marked states. At the $r^{th}$ iteration, the state of the algorithm is given by:
\begin{equation}
|\psi_r\rangle=G^r|\psi_0\rangle=\frac{\cos\theta_r}{\sqrt{2^n-M}}|S_0\rangle + \frac{\sin\theta_r}{\sqrt{M}}|S_1\rangle.
\label{Gro_rth}
\end{equation} 
Here, $|S_0\rangle$ is the superposition of the non-marked states, while $|S_1\rangle$ is the superposition of all the target states. At the $r^{th}$ iteration, $\theta_r= (r+\frac{1}{2})\sin^{-1}(2\sqrt{\frac{M}{N}})$ \cite{NieChu01, Gro96}. Clearly, in order to terminate in a superposition of the solution states, the optimal value of $r$ occurs when $\theta_r=\frac{\pi}{2}$. Thus $r_{opt}=CI[(\frac{\pi}{\sin^{-1}(2\sqrt{\frac{M}{N}})})-1)\frac{1}{2}]$, where $CI(x)$ returns the closest integer to $x$. Clearly, for $N>>M$, $r_{opt}=O(\sqrt{\frac{N}{M}})$.\\
 
The geometric measure of entanglement of a state $|\psi\rangle$ is expressed as its distance from its nearest separable state $|\zeta\rangle$. In other words, the overlap between $|\psi\rangle$ and $|\zeta\rangle$ is maximized and the entanglement of the state $|\psi\rangle$ is expressed as \cite{WeiGol08}
\begin{equation}
E(|\psi\rangle)=1-_{max_{\zeta}}|\langle\zeta|\psi\rangle|^2.
\label{Geom}
\end{equation} 
This value of entanglement $E(|\psi\rangle)$ of the state $|\psi\rangle$ can be thought of as the sine squared of its angle with its closest separable state $|\zeta\rangle$. This measure quantifies the amount of global entanglement that a quantum state inherits.\\
In this article, we have used the geometric measure of entanglement (\ref{Geom}) to analyse how the amount of entanglement varies across iterations in the Grover's search algorithm. Earlier, a similar approach had been taken in \cite{Rungta07} to quantify entanglement across each iteration of the Grover's Algorithm by using concurrence \cite{Woo98}. In \cite{Bru10}, the author reveals that the number of entangled states in the quantum search algorithm increase with the number of qubits.\\
The structure of the article is as follows. In Section II, the generalized expression for geometric measure of entanglement for $n$ qubits and $M$ solution states in the quantum search algorithm is derived.  In section III, explicit analytical relations of dependence of entanglement with the number of iterations and the maximum value of entanglement across all iterations are calculated. In section IV, the entanglement dynamics with the number of qubits $n$ and the number of solutions $M$ has been plotted using numerical methods. It also consists of entanglement dynamics when the quantum search algorithm converges to some fixed known states. Section V compares the geometric measure of entanglement with concurrence with respect to entanglement in the Grover's search algorithm. Finally, we conclude in section VI.
\end{section}
\begin{section}{Geometric measure of entanglement in the Grover's Search Algorithm}
As mentioned in Eq. (\ref{Gro_rth}) of Section I, the quantum state $|\psi_r\rangle$ at a given iteration $r$ of the Grover's algorithm is expressed as the superposition of the non-target and the target states. As the $r \rightarrow r_{opt}$, the amplitude of the solution states increase and that of the non-solution states decrease. Also, it is interesting to note that the Grover's iterate $G$ is comprised of two basic stages: First, phase inversion by an oracle and second, an inversion about the mean. It has been shown that at each iteration, it is the action of the oracle that leads to an increase in the amount of entanglement whereas, the second stage reduces the same \cite{Rungta07}.\\
Now, to compute the amount of entanglement at the $r^{th}$ iteration of the Grover's algorithm, let us assume a purely n-separable state $|\zeta\rangle=(\cos\frac{\phi}{2}|0\rangle+e^{i\gamma}\sin\frac{\phi}{2}|1\rangle)^{\otimes n}$. Each partition represents the most general form of a single qubit. The task is to maximize the overlap between $|\psi_r\rangle$ and $|\zeta\rangle$. For the algorithm, the value of $\theta_r$ lies in the interval $[0,\frac{\pi}{2}]$ (as shown in Eq. (1)) and this results in all coefficients of the state $|\psi_r\rangle$ to be positive. This enables us to fix $\gamma=0$,  when we are maximizing the overlap. Another, interesting observation is that the coefficients of $|\zeta\rangle$ are permutation invariant. This implies that the coefficients of all basis states in $|\zeta\rangle$ containing the same number of $0$'s and $1$'s are equal. Thus for all basis states with $n-k$ zeroes and $k$ ones, the coefficient is $\cos^{n-k}\frac{\phi}{2}\sin^k\frac{\phi}{2}$ and there are ${n}\choose{k}$ basis states having this coefficient.\\
Let us assume that of the $N=2^n$ states present in the database, only $M$ are solution states. Let us express the marked states in terms of the number of $0$'s and $1$'s they contain. Let the $M$ marked basis states contain $n_1,n_2,...n_m$ $0$'s respectively.\\
Making use of the phase optimality and permutation invariance of $|\zeta\rangle$, we arrive at its overlap with $|\psi_r\rangle$.
\begin{equation}
\begin{split}
\langle\zeta|\psi_r\rangle=\frac{\cos\theta_r}{\sqrt{N-M}}(\cos\frac{\phi}{2}+\sin\frac{\phi}{2})^n\\
+(\frac{\sin\theta_r}{\sqrt{M}}-\frac{\cos\theta_r}{\sqrt{N-M}})
(\sum_{i=1}^{M}\cos^{n-n_i}\frac{\phi}{2}\sin^{n_i}\frac{\phi}{2}).
\end{split}
\label{innerprod}
\end{equation}
Thus, entanglement of $|\psi_r\rangle$ is given by the following expression:
\begin{equation}
\begin{split}
E(|\psi_r\rangle)=1-_{max_{\phi}}|\frac{\cos\theta_r}{\sqrt{N-M}}(\cos\frac{\phi}{2}+\sin\frac{\phi}{2})^n+\\
(\frac{\sin\theta_r}{\sqrt{M}}-\frac{\cos\theta_r}{\sqrt{N-M}})(\sum_{i=1}^{M}\cos^{n-n_i}\frac{\phi}{2}\sin^{n_i}\frac{\phi}{2})|^2.
\end{split}
\label{entang_gen}
\end{equation}
Thus, for each iteration, we have obtained an expression that can quantify the entanglement. For various $n$ and $M$, we calculate the value of $E$ both analytically and numerically.
\end{section}
\begin{section}{Analytical results on dependence of entanglement on the number of iterations and maximum entanglement reached}
In this section, we establish analytically, a relation between the entanglement $E$ and the number of iterations $r$.
For the Grover's algorithm of $n$ qubits and $M$ solutions, the entanglement at the $r^{th}$ iteration is given by Eq. (\ref{entang_gen}).\\
Let, us assume $\phi=\phi _r$ to be the value of $\phi$ for which the overlap is maximum at the $r^{th}$ iteration, where $1\leq r\leq r_{opt}$. Thus the expression for entanglement at the $r^{th}$ iteration is as follows:
\begin{equation}
\begin{split}
E(|\psi_r\rangle)=1-(\frac{\cos\theta_r}{\sqrt{N-M}}(\cos\frac{\phi_r}{2}+\sin\frac{\phi_r}{2})^n+\\
(\frac{\sin\theta_r}{\sqrt{M}}-\frac{\cos\theta_r}{\sqrt{N-M}})(\sum_{i=1}^{M}\cos^{n-n_i}\frac{\phi_r}{2}\sin^{n_i}\frac{\phi_r}{2}))^2.
\end{split}
\label{ent_again}
\end{equation}
Assuming that $s_1=(\cos\frac{\phi_r}{2}+\sin\frac{\phi_r}{2})^n$ , $s_2=\sum_{i=1}^{M}\cos^{n-n_i}\frac{\phi_r}{2}\sin^{n_i}\frac{\phi_r}{2}$, $k_1=\frac{s_1-s_2}{\sqrt{N-M}}$ and $k_2=\frac{s_2}{\sqrt{M}}$, Eq. (\ref{ent_again}) gives:
\begin{equation}
E(|\psi_r\rangle)=1-(k_1\cos\theta_r+k_2\sin\theta_r)^2.
\end{equation}
Substituting $\lambda=\sqrt{1-E(|\psi_r\rangle)}$, we obtain a quadratic equation in $\cos\theta_r $. As $\theta_r$ is acute, we get:
%\begin{equation}
%\theta_\textbf{r}=\cos^{-1}(\frac{\lambda k_1^2 \pm k_2\sqrt{k_1^2+k_2^2-\lambda^2}}{k_1^2+k_2^2})
%\end{equation}
%where, $\theta_r=(r+1/2)\sin^{-1}(2\sqrt{\frac{M}{N}})$. This gives
\begin{equation}
r=\frac{\cos^{-1}(\frac{\lambda k_1^2 + k_2\sqrt{k_1^2+k_2^2-\lambda^2}}{k_1^2+k_2^2})}{\sin^{-1}(2\sqrt{\frac{M}{N}})}-1/2.
\label{r_vs_ent}
\end{equation}
As $\cos\theta_r$ is real, we have:
$k_1^2+k_2^2 \geq \lambda^2$.
Putting the values of $k_1, k_2$ and $\lambda$, we obtain a bound for the entanglement $E$ as:
\begin{equation}
E(|\psi_r\rangle)\leq 1-\frac{(s_1-s_2)^2}{N-M}-\frac{s_2^2}{M}.
\end{equation}
Thus we obtain the required expression. Here, the entanglement value never reaches $1$ as it can occur only when $s_1=s_2=0$. The maximum entanglement, $E_{max}$ across iterations ($1\leq r\leq r_{opt}$) is
\begin{equation}
E_{max}=max_r \{E(|\psi_r\rangle)\}.
\end{equation} 
\end{section}
\begin{section}{Quantifying entanglement: Numerical Results}
It becomes difficult to analytically obtain $\phi_r$ and quantify entanglement. Hence we resort to numerical analysis. Numerical results are obtained by considering $t=\tan\frac{\phi}{2}$ in Eq. (\ref{entang_gen}) and maximizing the resulting polynomial. Thus the problem is reduced to obtaining the roots of a polynomial as indicated in \cite{WeiGol08}. The entanglement value for each iteration $r$ is then plotted with the number of iterations varying from $r=1$ to $r=r_{opt}$. As the initial state of the Grover's algorithm is a result of an $n$ qubit Hadamard transform, the initial entanglement $E(|\psi_0\rangle)$ is always $0$.
\begin{subsection}{Entanglement when M=1}
When there exists a single marked state, the entanglement increases with the increase in the number of iterations, becomes maximum at exactly $\frac{r_{opt}}{2}$ and tails off to zero when $r=r_{opt}$. The nature of the curve is independent of the value of $n$ or the selection of $M$. The peak value of entanglement increases with the increase in $n$. The results for $n=7$ and $n=8$ qubits have been shown in Fig. \ref{onesol}, assuming that the state $|00..0\rangle$ is marked (result does not change on altering the marked state).
\begin{figure}[h]
\centering
\subfigure[Entanglement for n=7 qubits]{\includegraphics[scale=0.0335]{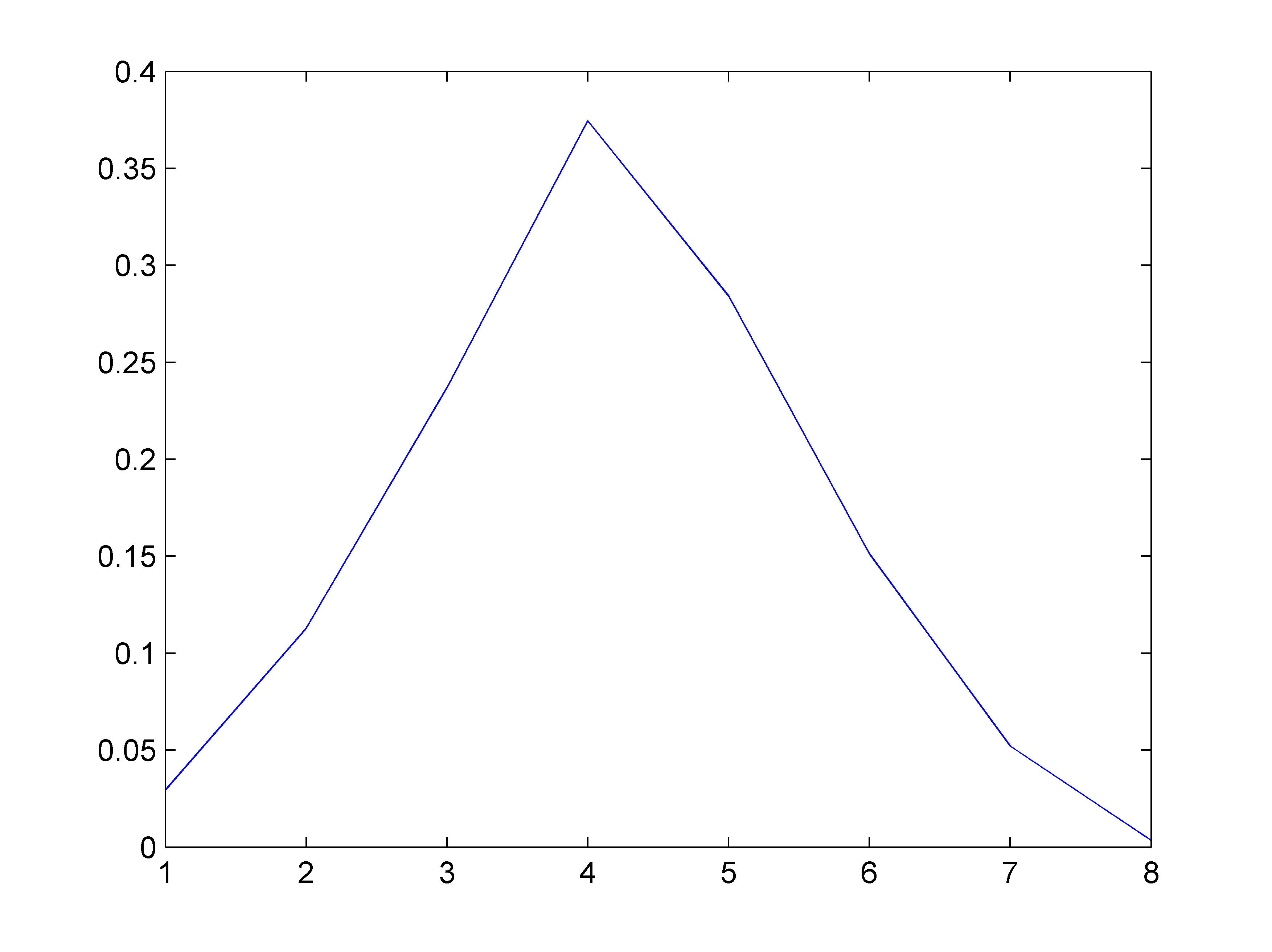}
}
\subfigure[Entanglement for n=8 qubits]{\includegraphics[scale=0.0335]{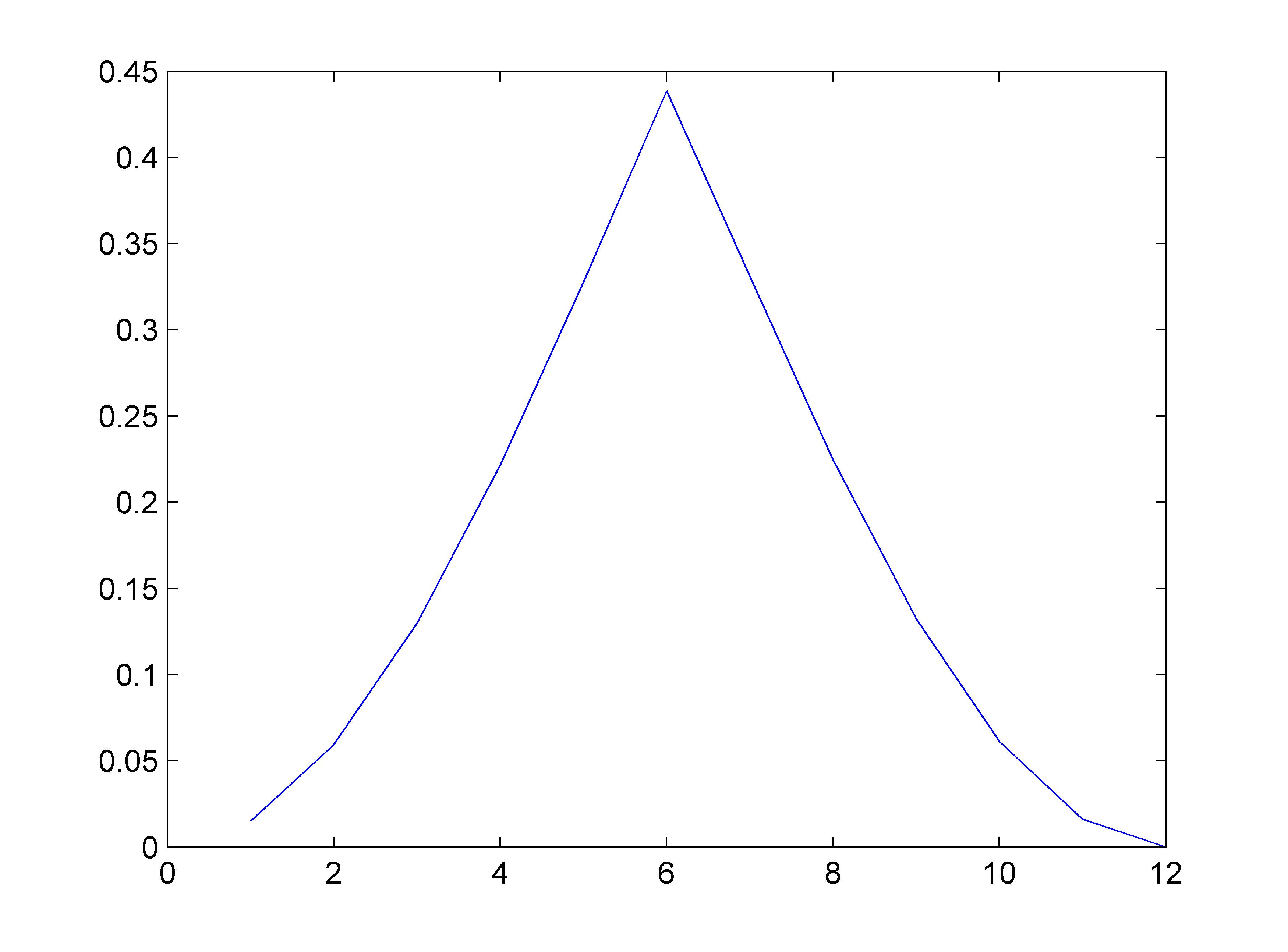}
}
\caption{{\small Entanglement with respect to the number of iterations when $M=1$ and $|00..0\rangle$ is marked. Here Y-axis depicts the entanglement while the number of iterations is shown along the X-axis.}}
\label{onesol}
\end{figure}   
For $n=7$ qubits, $r_{opt}=8$, and a peak entanglement value of $0.37$ is attained while the values for $n=8$ qubits are $12$ and $0.44$ respectively. This trend continues as $n$ increases. An interesting observation has been the fact that the peak value is attained at exactly half of the optimal number of iterations. This result adheres to the one found in \cite{Rungta07} using  $n$-qubit concurrence. The scenario changes for $M>1$. We analyse how the entanglement varies when $n$ is fixed and $M$ varies.
\end{subsection}
\begin{subsection}{Entanglement when n is fixed and M changes}
In order to study the variation of entanglement with the increase in the number of marked states, we fix the number of qubits fixed at $n=10$. Some of the results for $M=2$, $M=3$, $M=5$ and $M=10$ are shown in Fig. \ref{mfixed}. 
\begin{figure}[h]
\centering
\subfigure[Entanglement when M=2]{\includegraphics[scale=0.0335]{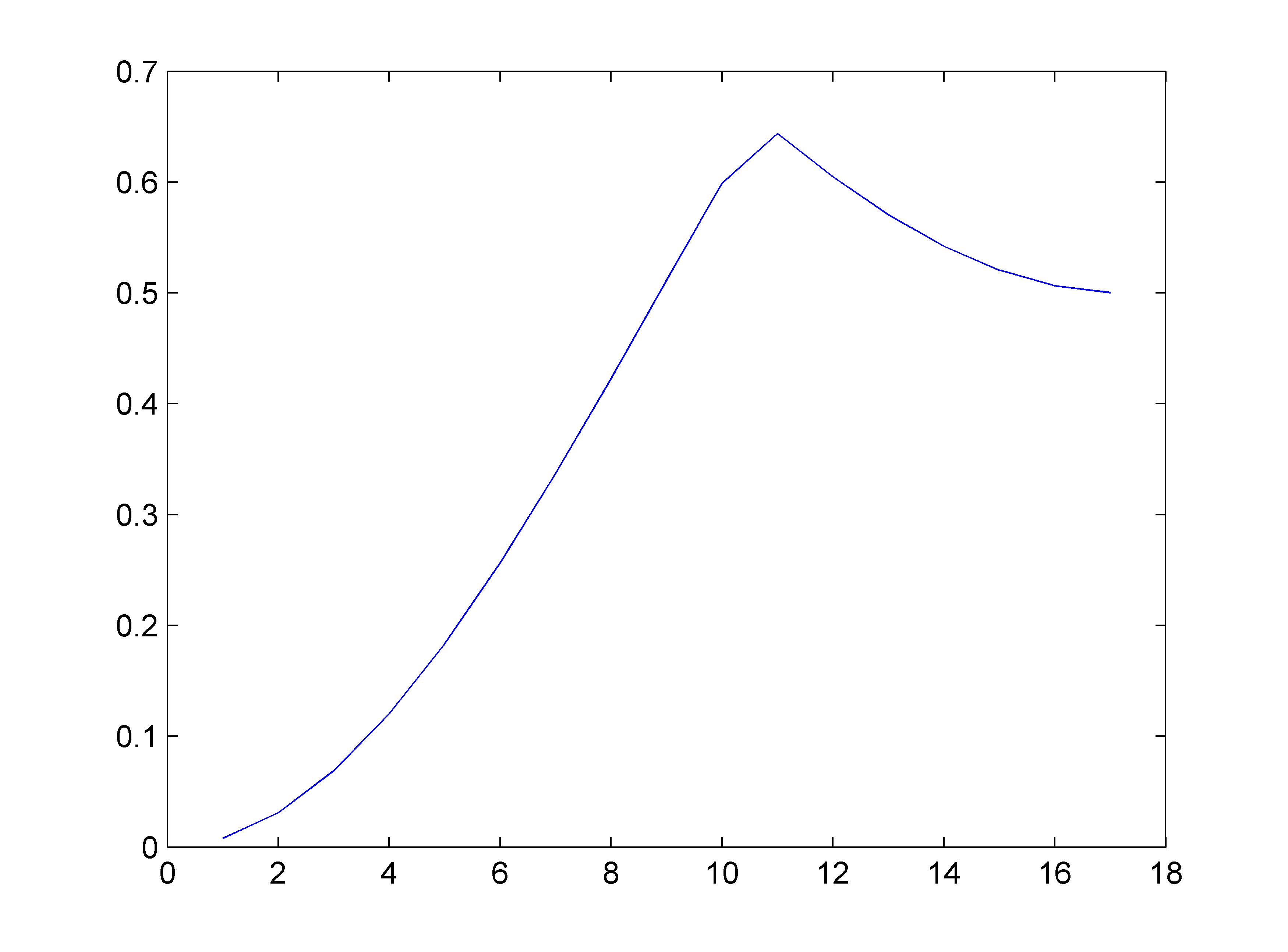} 
}
\subfigure[Entanglement when M=3]{\includegraphics[scale=0.0335]{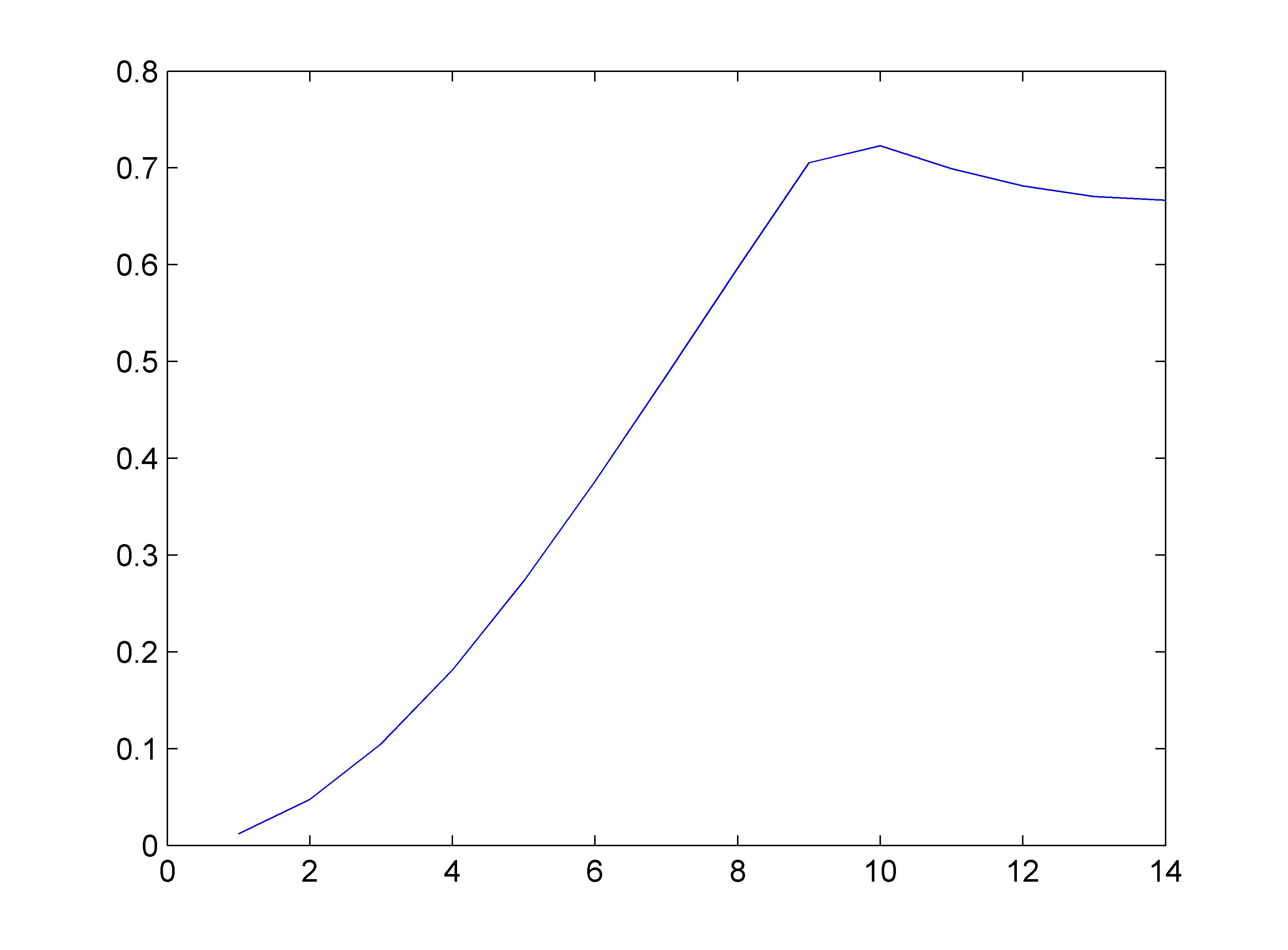}
}
\subfigure[Entanglement when M=5]{\includegraphics[scale=0.0335]{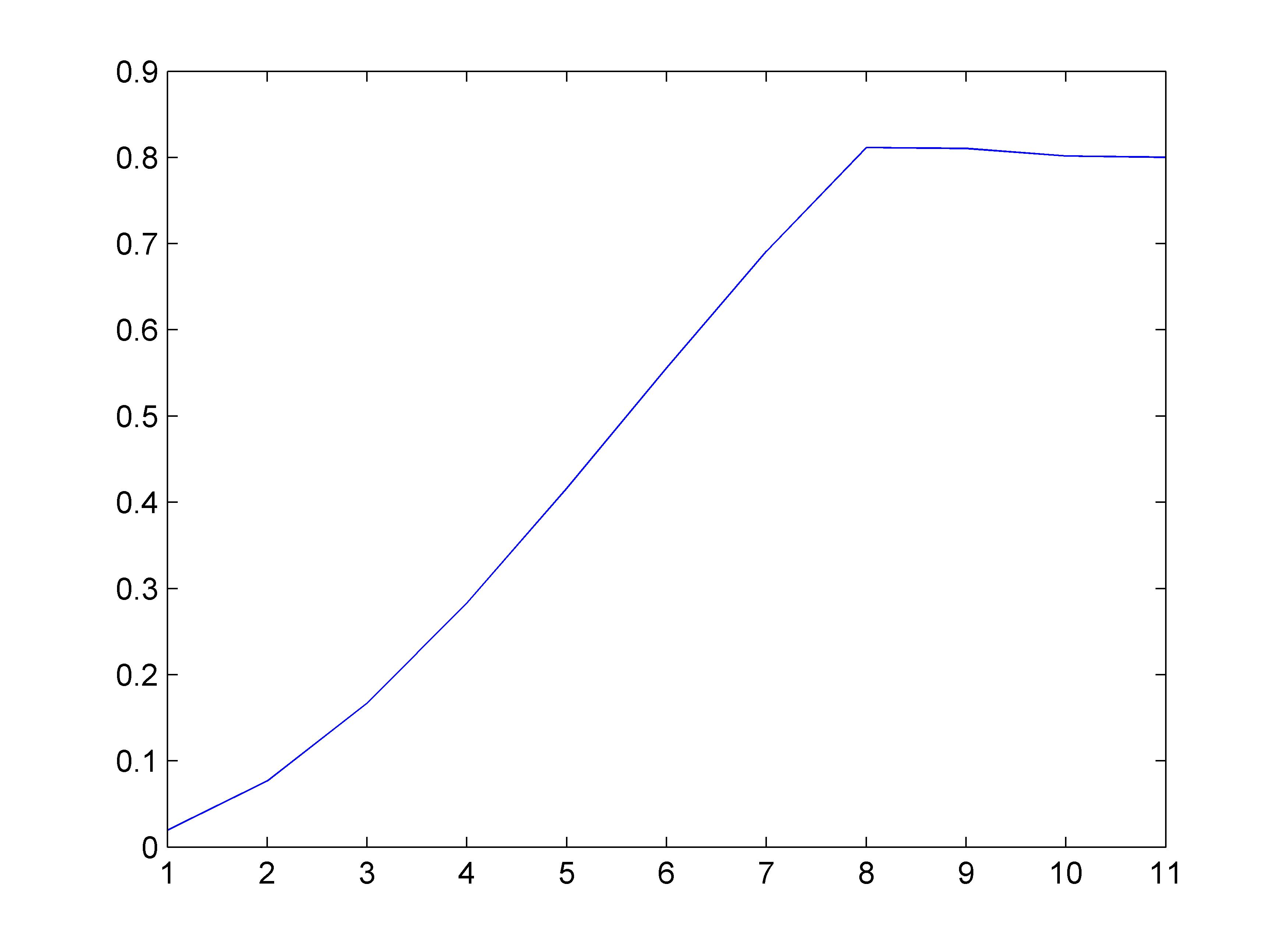}
}
\subfigure[Entanglement when M=10]{\includegraphics[scale=0.0335]{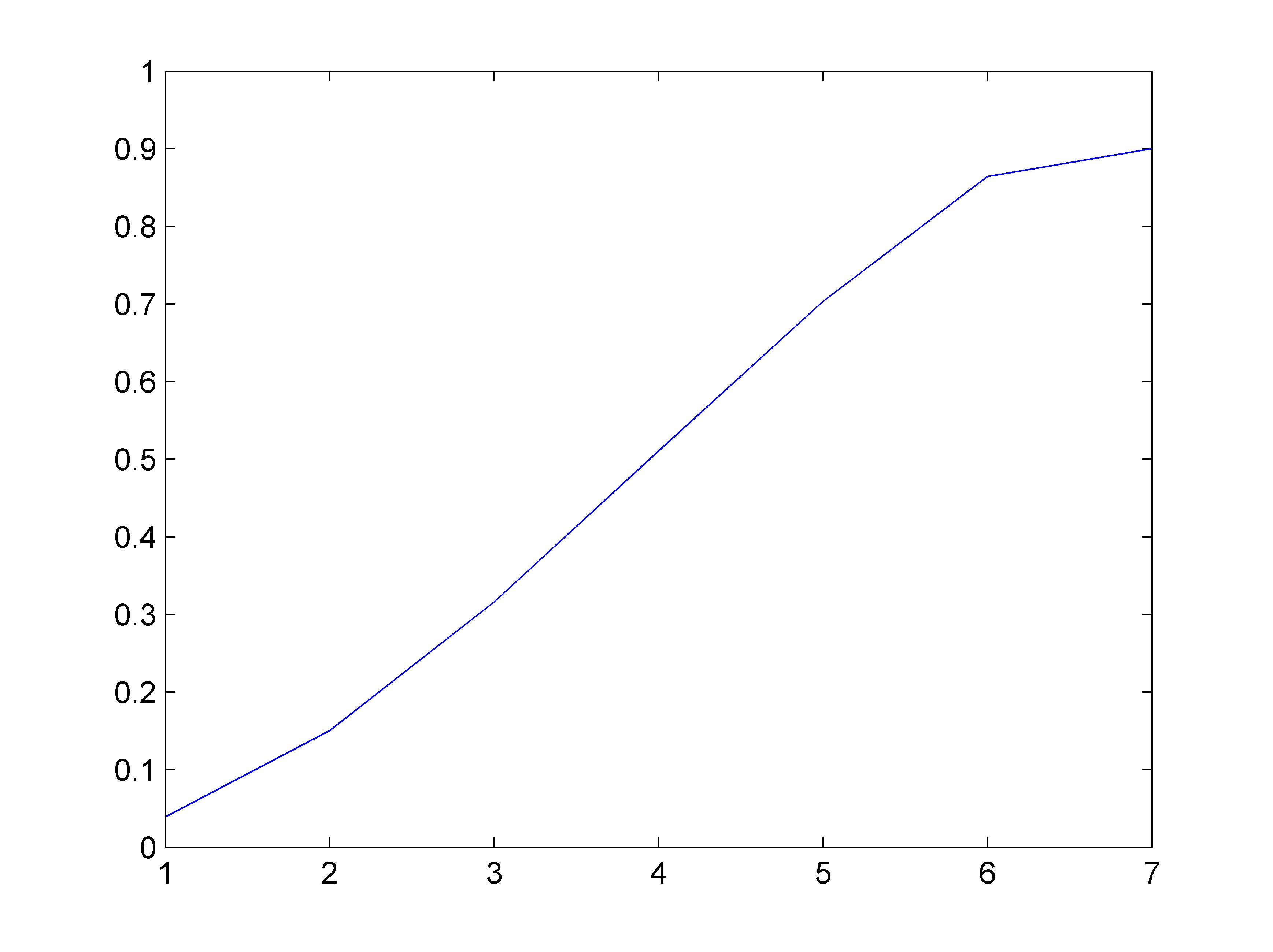}
}
\caption{{\small Entanglement with respect to the number of iterations when $n=10$ and $M$ changes. Here entanglement is plotted along the Y-axis while the number of iterations is plotted along the X-axis.}} 
\label{mfixed}
\end{figure}
Now, again the choice of marked states become important. As mentioned earlier, for $M=2$, $|00...0\rangle$ and $|11...1\rangle$ are chosen. For $M>2$, the states $|00..0\rangle$,$|11..1\rangle$ and $M-2$ states with $\frac{n}{2}$ 0's and $\frac{n}{2}$ 1's states are chosen as the marked states without any loss of generality. The entanglement for all such states is given by:
\begin{equation}
\begin{split}
E(|\psi_r,M\rangle) = 1-_{max_{\phi}}|\frac{\cos\theta_r}{\sqrt{N-M}}(\cos\frac{\phi}{2}+\sin\frac{\phi}{2})^n+\\(\frac{\sin\theta_r}{\sqrt{M}}-\frac{\cos\theta_r}{\sqrt{N-M}})(cos^n\frac{\phi}{2}+
\sin^{n}\frac{\phi}{2}+\\(M-2)\cos^{\frac{n}{2}}\frac{\phi}{2}\sin^{\frac{n}{2}}\frac{\phi}{2})|^2.
\end{split}
\end{equation}
We observe that the peak value of entanglement increases with increase in $M$. Also, interestingly, with an increase in $M$, the rise in entanglement decreases and it takes longer (more number of iterations)to reach the peak, or in other words, the peak shifts to the right. In the Fig. \ref{mfixed}, this has been exhibited clearly. Earlier, we had seen that for $M=1$, the entanglement peaked at exactly $0.5r_{opt}$, irrespective of the choice of the marked state.
\begin{table}[h]
\centering
\resizebox{5cm}{3cm}{
\begin{tabular}{|p{1.5cm}|p{1.5cm}|p{2cm}|}
\hline
 No. of Marked states   & Optimal no. of iterations  & No. of iterations required to reach peak entanglement\\
\hline
$1$  & $24$ & $0.5r_{opt}$\\
$2$ & $17$ & $0.647r_{opt}$\\
$3$ & $14$ & $0.714r_{opt}$\\
\vdots & \vdots & \vdots\\
$5$ & $11$ & $0.727r_{opt}$\\
\vdots & \vdots & \vdots\\
$10$ & $7$ & $r_{opt}$\\
\hline 
\end{tabular}
}
\caption{Iterations required to attain maximum entanglement for $n=10$ qubits}
\end{table}   
As shown in Table I, we find that the peak entanglement gets closer to the optimal number of iterations and ultimately coincides with the same. On increasing $n$, the value of $M$ required for maximum entanglement to coincide with $r_{opt}$ increases.
\end{subsection}
\begin{subsection}{Entanglement dynamics when the algorithm converges to physically known quantum states}
\begin{subsubsection}{GHZ state}
When $M=2$, the choice of the marked states become important and depending on this the expression and dynamics of entanglement changes. The entanglement value is $0$ in the beginning, increases with $r$ and attains a maximum value to the right of $\frac{r_{opt}}{2}$ and decreases therein till $r_{opt}$. The dynamics are unchanged with the selection of $M$. However, the peak entanglement value increases with an increase in $n$ just as the case with $M=1$. However, the final entanglement depends on the choice of the marked states. \\
The $n$ qubit GHZ state is defined as \cite{GHZ}
\begin{equation}
|GHZ\rangle_n = \frac{1}{\sqrt{2}}(|000...0\rangle + |111...1\rangle).
\end{equation}
When $|000..0\rangle$ and $|111...1\rangle$ are chosen as the marked states, the resulting final state is a GHZ state and the entanglement value is very close to $0.5$ as shown in Fig. \ref{ghzconv}. The expression for entanglement in that case is given by:
\begin{equation}
\begin{split}
E(|\psi_r,M=2\rangle) = 1-_{max_{\phi}}|\frac{\cos\theta_r}{\sqrt{N-2}}(\cos\frac{\phi}{2}+\sin\frac{\phi}{2})^n+\\(\frac{\sin\theta_r}{\sqrt{2}}-\frac{\cos\theta_r}{\sqrt{N-2}})(cos^n\frac{\phi}{2}+\sin^{n}\frac{\phi}{2})|^2.
\end{split}
\end{equation}
\begin{figure}[h]
\centering
\subfigure[Entanglement for n=7 qubits]{\includegraphics[scale=0.0335]{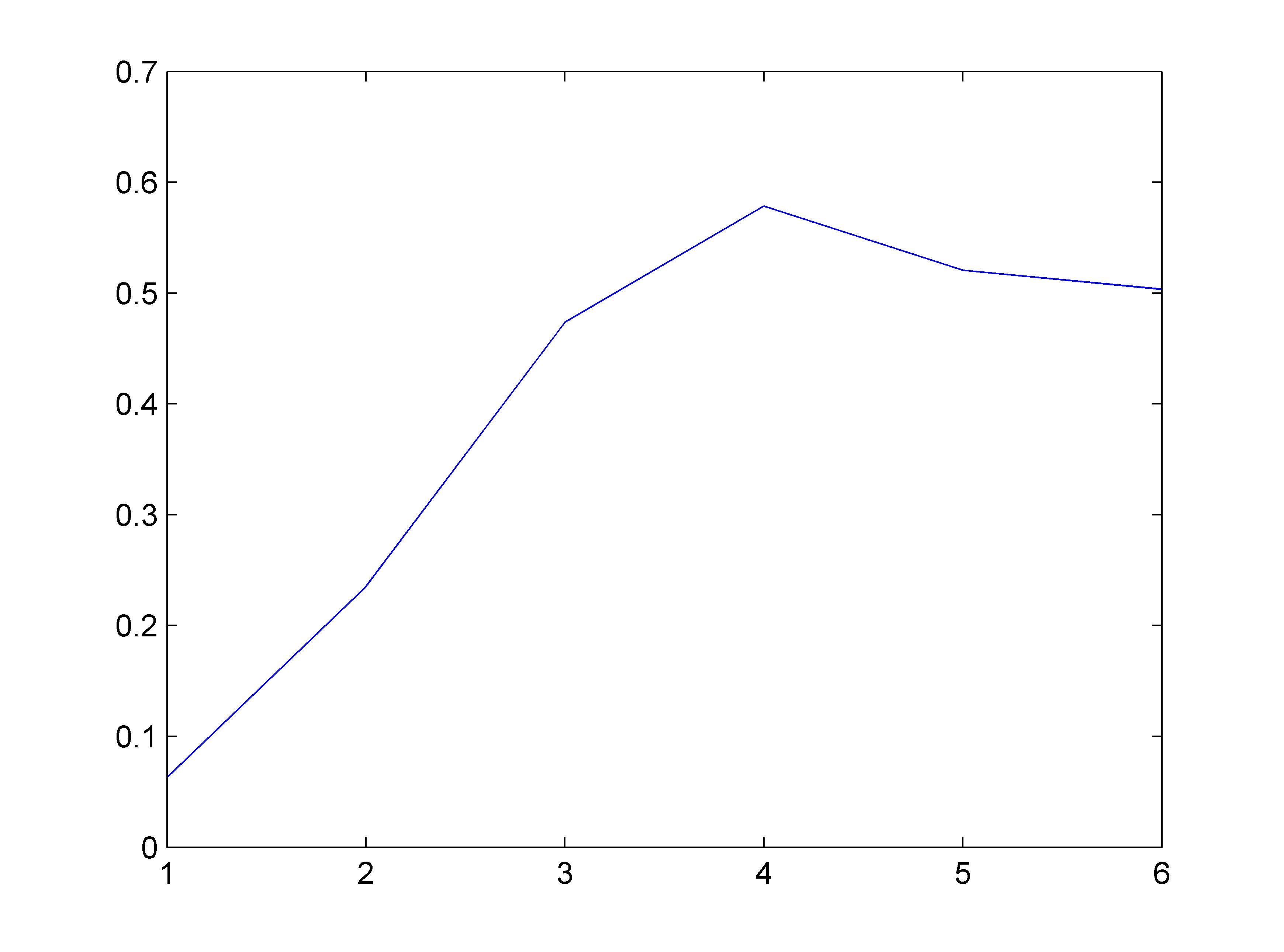} 
}
\subfigure[Entanglement for n=8 qubits]{\includegraphics[scale=0.0335]{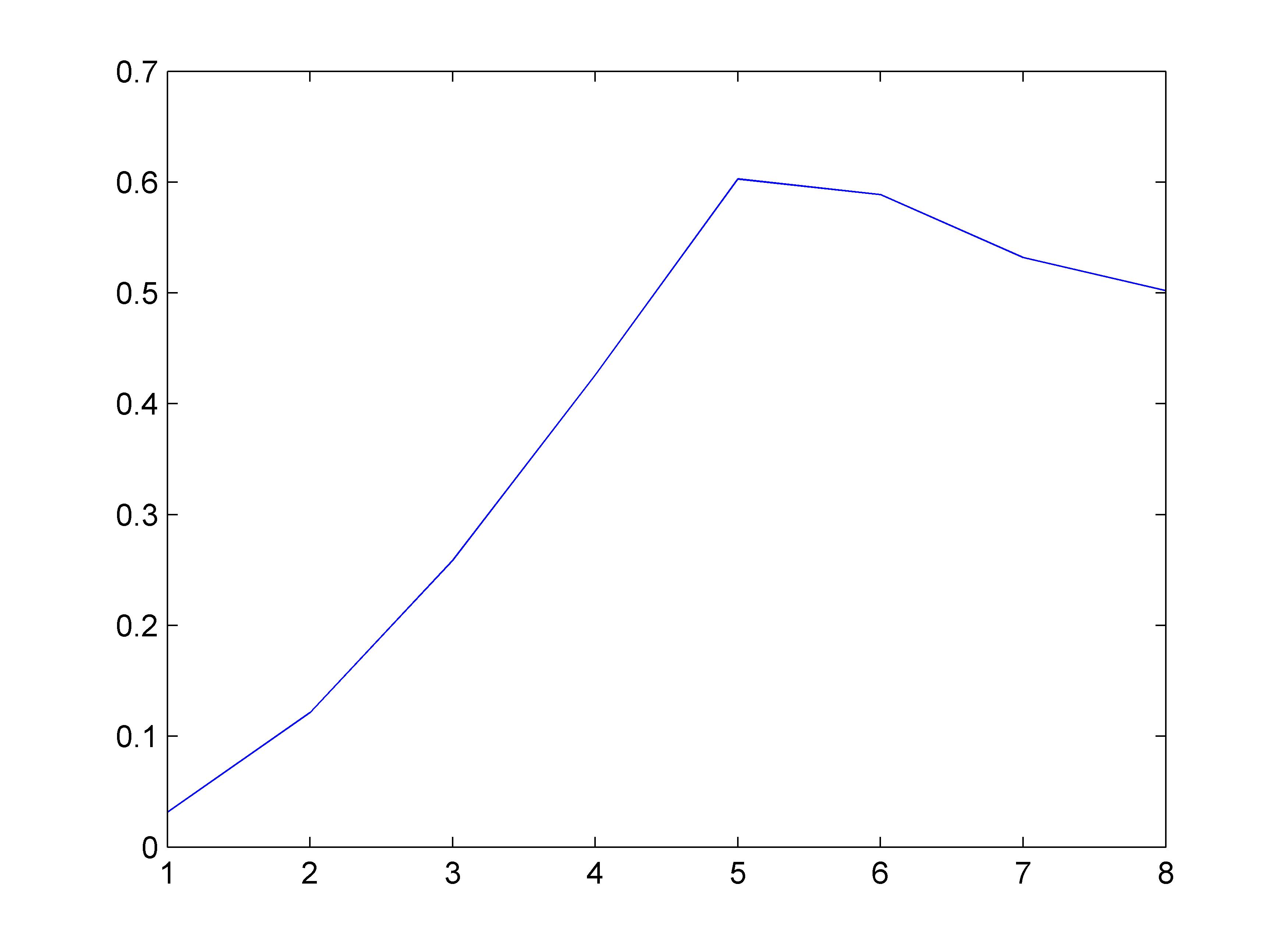}
}
\subfigure[Entanglement for n=9 qubits]{\includegraphics[scale=0.0335]{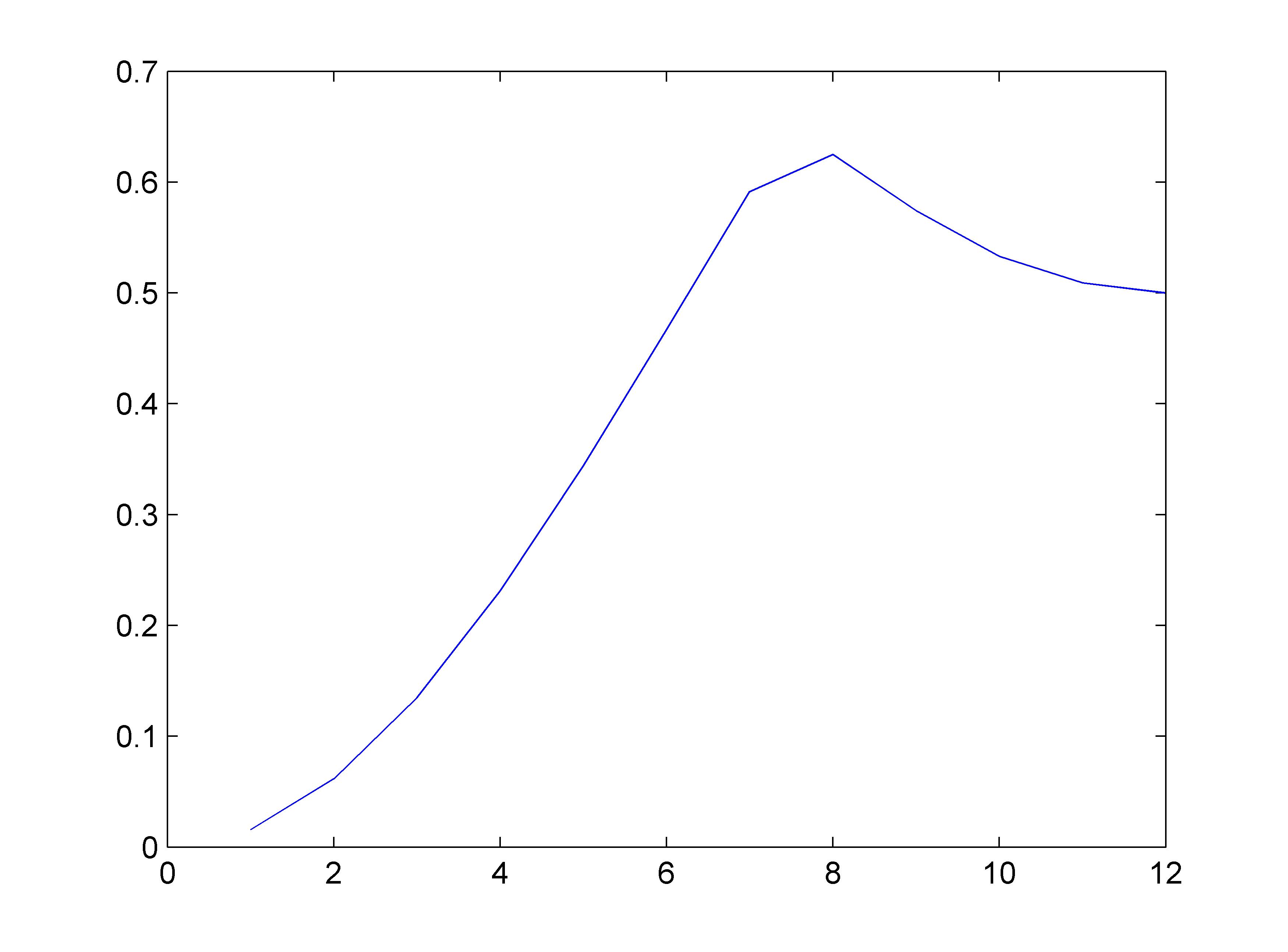}
}
\subfigure[Entanglement for n=10 qubits]{\includegraphics[scale=0.0335]{n=10_qubits_2.jpg}
}
\caption{{\small Entanglement dynamics with respect to the number of iterations when $|00..0\rangle$ and $|11..1\rangle$ are marked. Here Y-axis depicts entanglement and the number of iterations is shown in the X-axis.}} 
\label{ghzconv}
\end{figure}\\   
Clearly, although the nature of the curve remains the same, the maximum value of entanglement increases from $0.58$ to $0.64$ as $n$ changes from $7$ qubits to $10$ qubits. However, the entanglement of the final state is always $0.5$ and on changing the marked states, this value is altered.\\

Thus, for a fixed $M$ and on altering $n$, the dynamics of entanglement do not change. In the next section, we fix $n$ and alter $M$ and study the nature of the underlying entanglement.
\end{subsubsection}
\begin{subsubsection}{Dicke state}
In general, an $n$ qubit Dicke state \cite{Dicke} with $k$ excitations is given by:
\begin{equation}
|D_{k}^{n}\rangle=\frac{1}{\sqrt{\binom{n}{k}}}\sum_{j}\Pi_j\{|1\rangle ^{\otimes k} |0\rangle ^{\otimes n-k}\},
\end{equation}
where $\sum_j \Pi_j$ denotes the sum over all possible permutations of $n-k$ $0$'s and $k$ $1$'s. These states are well known in quantum optics and have appeared in a number of investigations related to the phenomena of superradiance \cite{Dicke, Prasad}.\\
 As all the amplitudes of $|D_{k}^{n}\rangle$ are positive, the nearest separable state $|\zeta\rangle$ would be:
\begin{equation}
|\zeta\rangle=(\cos\frac{\phi}{2}|0\rangle + \sin\frac{\phi}{2}|1\rangle)^{\otimes n}.
\end{equation}
Thus, 
\begin{equation}
\langle\zeta|D_{k}^{n}\rangle = \sqrt{\binom{n}{k}}\cos^{n-k}\frac{\phi}{2}\sin^k\frac{\phi}{2}
\end{equation}
and the geometric measure of entanglement is given by
\begin{equation}
E(|D_{k}^{n}\rangle)=1-max_{\phi}|\sqrt{\binom{n}{k}}\cos^{n-k}\frac{\phi}{2}\sin^k\frac{\phi}{2}|^2.
\end{equation}  
By converting the above equation into a polynomial and maximizing it over $\phi$ yields the value of entanglement as
\begin{equation}
E(|D_{k}^{n}\rangle)=1-\{\binom{n}{k} \frac{k^k}{n^n}(n-k)^{(n-k)}\}.
\end{equation}
Thus by varying $k$ we can obtain a plethora of Dicke states. One such example is $|D_{1}^{n}\rangle$ which is the generalized $n$ qubit $W$ state \cite{Berg13}. The Grover's algorithm converges to the $W$ state if the marked states are aptly chosen.
\begin{subsubsection}{W state}
The generalized n-qubit W state is a maximally entangled state \cite{W} and is expressed as:
\[|W_n\rangle=\frac{1}{\sqrt{n}}(|100...0\rangle+|010...0\rangle+...+|000...1\rangle).\]
The maximum overlap between $|W_n\rangle$ and $|\zeta\rangle$ is calculated as:
\[_{max}|\langle\zeta|W_n\rangle|=\sqrt{\frac{n}{n-1}}(\frac{n-1}{n})^{n/2}=(\frac{n-1}{n})^{(\frac{n-1}{2})}\]
Thus, 
\begin{equation}
E(|W_n\rangle)=1-(\frac{n-1}{n})^{(n-1).}
\end{equation}  
Clearly, the entanglement value of W states is greater than that of GHZ states. This occurs because the geometric measure of a quantum state is calculated from its nearest \textit{n separable} state and is a global entanglement measure, not quantifying genuine multipartite entanglement. To quantify genuine multipartite entanglement of a state, its overlap from its nearest bi-separable state must be calculated \cite{Hier08}. In this section, we analyse the dynamics of entanglement when the algorithm converges to a W state.\\
\begin{figure}[h]
\centering
\includegraphics[scale=0.04]{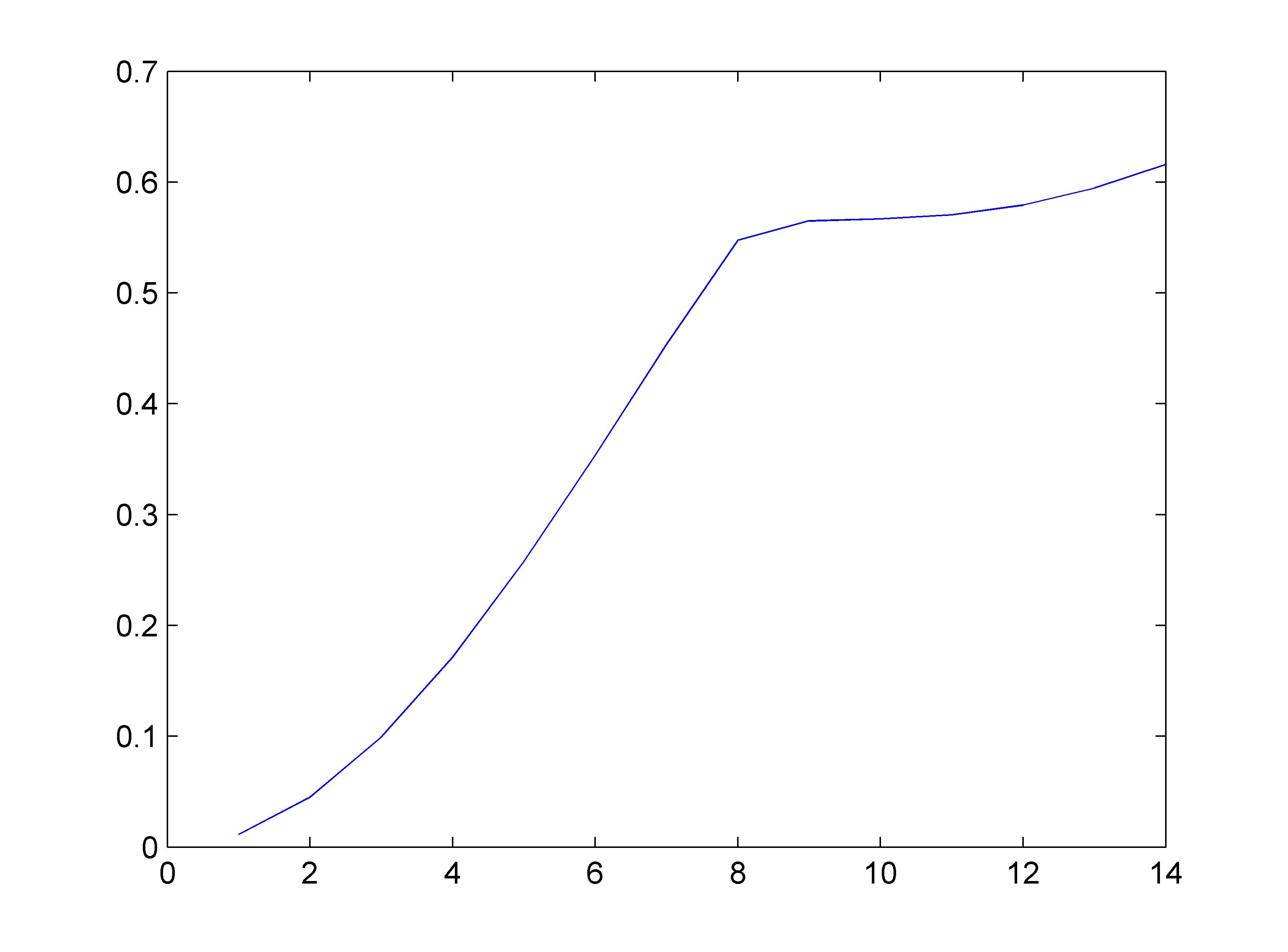}
\caption{{\small Entanglement dynamics with respect to the number of iterations when $n=12$ and the target state is a W state. Here entanglement is plotted along the Y-axis and the number of iterations along the X-axis.}} 
\label{wstate}
\end{figure} 
When the marked states $M=n$ and each such marked basis state contains exactly one $1$. The expression for geometric measure of entanglement at each iteration of the algorithm in such a case is given by:
\begin{equation}
\begin{split}
E(|\psi_r,M=n\rangle) = 1-_{max_{\phi}}|\frac{\cos\theta_r}{\sqrt{N-M}}(\cos\frac{\phi}{2}+\sin\frac{\phi}{2})^n+\\(\frac{\sin\theta_r}{\sqrt{M}}-\frac{\cos\theta_r}{\sqrt{N-M}})(n\cos^{n-1}\frac{\phi}{2}\sin\frac{\phi}{2})|^2.
\end{split}
\end{equation}
The entanglement dynamics for $n=12$ qubits is shown in Fig \ref{wstate}.
\end{subsubsection}
\end{subsubsection}
\end{subsection}
\end{section}
\begin{section}{Comparison with concurrence}
In \cite{Rungta07}, concurrence was used to quantify the entanglement at each iteration of the algorithm. The concurrence at the $r^{th}$ iteration was expressed in terms of the change in probability of obtaining the target state with respect to the number of iterations.
\[C(|\psi_r)\rangle=\frac{1}{2A_0}\frac{dA_r^2}{dr}.\]
Here, $A_r^2$ is the probability of obtaining the target state and $A_0$ is the initial amplitude of the superposition of marked states. For the case where there exists only one marked state, the evolution of concurrence with respect to the number of iterations follows a trend that is similar to the one obtained in the case of geometric measure of entanglement. The concurrence of the initial state $C(|\psi_0\rangle)$ and that of the final state, after $r_{opt}$ number of iterations is $0$. Other than that, concurrence is non-zero for all values of $r$.\\
On the other hand, for multiple marked states, the presence of entangled states was indicated without explicitly quantifying the same. Geometric measure of entanglement allows us to quantify entanglement for the presence of one or more marked states. Moreover, the expression for entanglement in our study allows us to analyse the variation in entanglement with increase in the number of qubits and also with the change in the number of marked states. 
\end{section}
\begin{section}{Conclusion}
In this article, we have studied the nature of entanglement in the Grover's search algorithm. At each iteration of the algorithm, the amount of entanglement has been precisely quantified using the geometric measure of the entanglement. As mentioned earlier, this entanglement value is a global entanglement quantifier and is not a measure of the inter-particle entanglement, i.e., it does not quantify genuine multipartite entanglement. In order to calculate the genuine multipartite entanglement, one needs to calculate the overlap of a state from its nearest bi-separable state with the bi-partition occurring between the first qubit and the remaining $n-1$ qubits. \\
A generalized expression for the entanglement in the Grover's algorithm for $n$ qubits and $M$ solution states has been calculated. This has been used to analyse the variation of entanglement with change in $n$ and $M$. The generic nature of the behaviour of entanglement does not alter with increase in $n$ for a given $M$. However, the maximum value of entanglement increases gradually. The amount of entanglement in the final state depends solely on the choice of the solution states as the algorithm ultimately terminates in an equal superposition of the target states. For $M=1$, the entanglement tails off to zero as the algorithm stops or the optimal number of iterations is reached, as the state is fully separable. Also, the maximum value of entanglement is reached at exactly half of the optimal number of iterations.\\
For $M>1$, the peak value of entanglement is no longer at the center but is shifted to the right. The choice of marked states, may lead to the termination of the algorithm in a GHZ or a W state. For a given value of $M$, the peak entanglement increases with $n$ and the position of the peak is same for all $M$. However, when $n$ is fixed and $M$ is increased gradually, the peak value of entanglement shifts gradually to the right and converges to a steady value at $r_{opt}$ number of iterations.\\
The dependence of entanglement on the number of iterations is calculated analytically which further imposes a bound on the amount of entanglement that can be attained during the course of the algorithm. We have also compared our results with that of multi-qubit concurrence. In \cite{Meyer2}, a global entanglement measure was defined and used to describe the evolution of entanglement in the Grover's search algorithm for ten qubits and one marked state. The entanglement dynamics is seen to be consistent with our results using geometric measure of entanglement. \\
%In future, it would be interesting to examine the nature of entanglement beyond $r_{opt}$. Although the algorithm might end, the evolution of entanglement beyond $r_{opt}$ might offer useful insights. It can be conjectured that for a sufficient number of iterations, the entanglement dynamics would always be the same--the one that is observed when $M=1$. For all $M$ and $n$, there exists a value of the number of iteration $r \geq r_{opt}$ such that entanglement falls to zero.  
\end{section}

\end{document}